\documentclass{article}
\usepackage{arxiv}
\usepackage{url}            % simple URL typesetting
\usepackage[utf8]{inputenc} % allow utf-8 input
\usepackage[T1]{fontenc}    % use 8-bit T1 fonts
\usepackage{hyperref}       % hyperlinks
\usepackage{booktabs}       % professional-quality tables
\usepackage{amsfonts}       % blackboard math symbols
\usepackage{nicefrac}       % compact symbols for 1/2, etc.
\usepackage{microtype}      % microtypography
\usepackage{lipsum}		% Can be removed after putting your text content
\usepackage{graphicx}
\usepackage{doi}

\usepackage{enumitem}
\usepackage{algorithmic}
\usepackage{comment}
\usepackage{url}
\usepackage{subfig}
\usepackage{float}
\usepackage[normalem]{ulem}
\useunder{\uline}{\ul}{}
\usepackage{booktabs}

\usepackage{graphicx}
\usepackage{color}
\usepackage{xcolor}
\usepackage{tipa}
\usepackage{amsmath}
\usepackage{csquotes}
\usepackage{tcolorbox}

\definecolor{RED}{RGB}{250,70,22}
\definecolor{seafoam}{RGB}{78,255,172}
\begin{document}

\title{Analyzing the AI Nudification Application
Ecosystem
}

\author{{\rm Cassidy Gibson}\\ University of Florida
\and
{\rm Daniel Olszewski}\\
University of Florida
\and
{\rm Natalie Grace Brigham}\\ University of Washington
\and
{\rm Anna Crowder }\\
University of Florida \vspace{0.25cm}
\and 
{\rm \hspace*{2.5cm} Kevin R. B. Butler} \\ \hspace*{2.5cm} University of Florida
\and
{\rm Patrick Traynor \hspace*{2.5cm} }\\ University of Florida  \hspace*{2.5cm} \vspace{0.25cm} 
\and
{\rm \hspace*{2.5cm} Elissa M. Redmiles}\\ \hspace*{2.5cm} Georgetown University 
\and
{\rm Tadayoshi Kohno \hspace*{2.5cm}  }\\University of Washington \hspace*{2.5cm}}

\pagestyle{plain}

\maketitle

\begin{abstract}
\label{sec:abstract}

Given a source image of a clothed person (an image subject), AI-based nudification applications can produce nude (undressed) images of that person. Moreover, not only do such applications exist, but there is ample evidence of the use of such applications in the real world and without the consent of an image subject. Still, despite the growing awareness of the existence of such applications and their potential to violate the rights of image subjects and cause downstream harms, there has been no systematic study of the nudification application ecosystem across multiple applications. We conduct such a study here, focusing on 20 popular and easy-to-find nudification websites. We study the positioning of these web applications (e.g., finding that most sites explicitly target the nudification of women, not all people), the features that they advertise (e.g., ranging from undressing-in-place to the rendering of image subjects in sexual positions, as well as differing user-privacy options), and their underlying monetization infrastructure (e.g., credit cards and cryptocurrencies). We believe this work will empower future, data-informed conversations — within the scientific, technical, and policy communities — on how to better protect individuals’ rights and minimize harm in the face of modern (and future) AI-based nudification applications.

\textcolor{red}{Content warning: This paper includes descriptions of web applications that can be used to create synthetic non-consensual explicit AI-created imagery (SNEACI). This paper also includes an artistic rendering of a user interface for such an application.} 

\end{abstract}

\section{Introduction}
\label{sec:intro}

Computer vision and generative AI techniques can undress\footnote{We define undress functionality as that which takes a representation (image) depicting a clothed individual and produces an representation of that individual without clothes.} someone depicted in a picture or video~\cite{yu2018generative, luo2020full,huang2022review,bualan2008naked}. 
Such technology is increasingly publicly accessible, leading to a proliferation of ``nudification'' applications available to end-users online. 

Nudification applications enable end-users without technical or even Photoshop skills to artificially generate intimate imagery of someone \emph{without} their consent. We refer to such a resulting image as a \textit{synthetic non-consensual explicit AI-created imagery}, or \textit{SNEACI}\footnote{Synthetic non-consensual explicit AI-Created imagery, or SNEACI, refers to images or videos that depict a nude or semi-nude subject, including those that contain intimate body parts and/or depict the subject engaged in a sexual act, without the subject's consent.}. Non-consensual imagery created through the use of AI has a starkly different paradigm compared to that of images created through Photoshop or technical skills. AI makes the creation of these images easier, faster, and more realistic because of how advanced generative AI has become. Similar to malware-as-a-service, synthetic non-consensual explicit AI-created imagery brings non-experts the ability to harm at scale. 
The creation of SNEACI is a form of sexual abuse against the subject who, by definition, is non-consensually depicted in the resulting content~\cite{mcglynn2017beyond}. In addition to the serious mental health impacts victim-survivors sustain from the violation of being depicted in SNEACI~\cite{henry2020image}, 
SNEACI may be used by the creator to extort and/or otherwise harass the subject of the image~\cite{marchal2024generative}. As such, there is increasing concern amongst policymakers and new legislation about SNEACI and the applications that facilitate its creation~\cite{UK:gov:law,who-cta,ExecutiveOrder}.
Despite the severity of abuse possible with nudification applications, little is known about this software ecosystem. In order to secure potential victims of an abusive software ecosystem, we must first understand how that ecosystem operates in practice~\cite{anderson1993cryptosystems}. We take a first step toward filling this gap through measuring and characterizing the ecosystem by answering  three key research questions. 

Our first research question is:
\begin{itemize}
    \item \textbf{RQ1.} How do nudification applications position themselves to clients via text and visual descriptions?
\end{itemize}
Acknowledging that nudification applications could be used with the consent of the image subject, we seek to understand whether existing applications foster and support those (and only those) use cases.
Additionally, we seek to understand the general experience of users as they interact with these sites. Among the questions we ask: do users need to confirm that they are adults, do the sites support the nudification of all people,
and how do the sites communicate about the role of consent from the image subject? Understanding the user experience on these sites, and how these sites position themselves and communicate with their users, is essential for having an informed conversation about their dominant use cases, the benefits and harms of these applications, and, to the extent that harms dominate benefits, how the broader research, industry, and policy communities might go about mitigating those harms. 

Next, we seek to answer the following research question:
\begin{itemize}
    \item \textbf{RQ2.} What features do nudification applications advertise?
\end{itemize}
In our study, we seek to understand and catalogue the advertised features of popular nudification applications --- these are the features that they purport to offer, and hence the features that users believe that the applications will provide. We chose not to experimentally verify that each nudification site actually provides all (and has no hidden features) the features that they claim, nor did we upload images of people for nudification to these sites to test these features, as we did not believe it would be ethical to upload images of real people to potentially adversarial entities, including images of people on our research team, or even already-public stock photos or images of celebrities. Instead, for our purposes, we focus on what features users believe they will have access to if they choose to use the nudification site. Even if not all nudification platforms provide all the features that they claim to, if users seek those features today, we conjecture that future instantiations of these platforms will provide those features. Hence, knowing the features that these applications advertise, and what features these applications believe that their users want or will want, is valuable.

Lastly, we seek to answer the following research question:
\begin{itemize}
    \item \textbf{RQ3.} How do nudification applications monetize?
\end{itemize}
Prior to the full initiation of our research study, we gained preliminary experience through the interaction with several nudificaiton sites. Our preliminary interactions uncovered a diverse monetization  ecosystem, including both conventional payment systems (like credit cards) as well as newer, less conventional payment systems (like cryptocurrencies). Under the hypothesis that at least some nudification sites are problematic, one approach toward curbing their existence might be to challenge their ability to monetize. Additionally, as part of studying the monetization  ecosystem, we sought to assess these sites' monetization  strategies, e.g., do certain features (like full nudification) require a paid subscription whereas other features (like changing clothing) do not?

To answer our research questions, we collected a sample of 20 nudification applications (websites) and systematically analyzed them using the application walkthrough method~\cite{light2018walkthrough} --- a methodology used in several fields within and beyond computer science (e.g., \cite{mattinen2023ruse,reime2023walking,o2024unpacking,wang2021protection,jarrahi2020platformic,duguay2017dressing}).

In analyzing the applications, we found a problematic ecosystem:
\begin{itemize}
    \item 19 out of 20 applications explicitly specialize in the undressing of women; only half of the websites mention that they expect the user to have the image subject's consent and fewer ask for affirmation that consent has been obtained.
    \item Most of the applications allowed for additional features beyond ``undressing'' (e.g., making the image-subject nude with their breasts and vulva visible in the imagery). For example, half of the applications allowed users to put image-subjects into sexual acts.
    \item These nudification applications make up a commercial ecosystem and, hence, targeting their commercialization features might be one way to protect against SNEACI.
    Furthermore, we see purposeful repackaging of these nudification features with 5 out of the 20 applications offering API access to their highest paying customers. 
\end{itemize}

Stepping back, the computer security research community, as it is often defined, focuses on computing in the presence of adversaries. In some cases, the research focuses on studying adversarial features and the adversarial ecosystem, and in other cases, the research focuses on studying defenses against said adversaries. The ecosystem of image-based sexual abuse is an example of an adversarial usage of technology, and this work sits within that context and builds on prior research in and adjacent to the computer security research community on understanding and studying the emerging realm of synthetic explicit non-consensual AI-created imagery~\cite{BrighamNCEI, mirandaIBSA, umbachNCEI}.

\section{Background and Related Work on SNEACI}
\label{sec:relwork}
In this section, we provide broader context and historical information regarding the rise of synthetic non-consensual explicit AI-created imagery (SNEACI) as well as related scholarly works. We discuss other related work inline in the body of this paper. 

Media ``deepfake''s\footnote{``Deepfake'' is the widely used colloquial terminology for media generated through AI. However, this terminology can often have a negative connotation because of its historical use in politics, especially against women.}~(i.e., image, video, etc) have become more widespread in the last decade both in their frequency and uses~\cite{fakenews, faceswapping}. 

The most common response to stopping the dissemination of these deepfakes is through the creation of deepfake detectors. Prior work has developed machine learning models trained to detect if the video is fake or different methods to spot defects in the video to determine that they are generated media~\cite{videodetection}. 
However, deepfake non-consensual explicit imagery (SNEACI) needs to be addressed as a form of sociotechnical violence, where detecting the output of that violence is not sufficient to remedy it. %in different ways than pure detection. 

SNEACI is a social issue of sexual harassment and abuse being perpetrated through  deepfake models, largely against women~\cite{pornstudies}. 
Deepfake generators have been used to create SNEACI since 2017 and often were centralized around women 
politicians~\cite{pornstudies}. These political deepfake SNEACI's have even been seen as recently as 2024 against a woman politician in Miami\footnote{This person has since dropped from the race that she was in.} and have also recently been seen used against celebrities like Taylor Swift and social media content creators like Pokimane~\cite{floridancei,pokimanencei,tswiftncei}. 

In the past, individuals have encountered image-based sexual abuse (IBSA) where real explicit images were spread across the internet or to specific parties to extort, exact revenge, or control a victim
~\cite{mirandaIBSA}. Since then, laws have been put in place in 47 of the 50 states in the United States of America which have criminalized IBSA in varying degrees~\cite{lawIBSA}, though such activities continue.
SNEACI creates a form of IBSA where the images are ``deepfakes'', or synthetically generated imagery.

New research within the computer security and adjacent communities have begun to evaluate awareness, prevalence, and societal norms around the creation, sharing and viewing of this content ~\cite{BrighamNCEI, umbachNCEI}. Additionally, research has explored how those targeted by SNEACI might seek help~\cite{mirandaIBSA}. Prior work has also studied perpetrator discourse~\cite{deepfakeComm}, examining how deepfake creator communities discuss the technology~(both in positive conversation and malicious use-cases) and how past attempts at mitigating abusive use cases have only caused the community to distrust moderation more~\cite{deepfakeComm}. In sum, past work on deepfakes primarily focus on human perspectives~\cite{brieger2024empowerment} --- about the technologies they use to create deepfakes, victim-survivor experiences, or societal perceptions --- rather than studying the systems capable of producing SNEACI. We seek to address the latter in this work through measuring and characterizing the AI nudification ecosystem. Such examination is necessary to understand how these technologies may be problematic and to think about how to defend against such abusive applications (e.g., intervene against their monetization sources, technical infrastructure, etc.) as has been done in other areas such as Stalkerware~\cite{gibson2022}. 

\section{Methodology}
\label{sec:methodology}

Nudification applications have yet to be characterized. These applications are not only technical --- their features build on computer vision and AI techniques --- but sociotechnical: to the extent that they are used without the consent of the person in an image, their use is embedded in the sociocultural context of sexual abuse. Thus, we adopt a method to examine this ecosystem through a sociotechnical lens: 
Light et al.'s application walkthrough method~\cite{light2018walkthrough}. This method expands on the HCI method of cognitive walkthroughs~\cite{wharton1994cognitive} used to assess issues of usability -- which was previously applied in the security literature as early as 1999 in ``Why Johnny Can't Encrypt''~\cite{whitten1999johnny} --- to examine an app's interface in order to assess:
\begin{displayquote}
    \textit{``its technological mechanisms and embedded cultural references to understand how it guides users and shapes their experiences. The core of this method involves the step-by-step observation and documentation of an app's screens, features and flows of activity --- slowing down the mundane actions and interactions that form part of normal app use in order to make them salient and therefore available for critical analysis \ldots this process is contextualised within a review of the app's vision, operating model and governance.~\cite{wharton1994cognitive}''}
\end{displayquote}
In this section, we both review the Light et al. method~\cite{light2018walkthrough} and detail our approach to selecting a representative sample of applications, collecting the data, and analyzing that data via a walkthrough analysis. We additionally discuss ethical considerations, our positionality, and limitations of our approach. 

\subsection{Dataset}
\label{sec:methodology:dataset}
We leveraged two categories of sources for identifying AI nudification applications, i.e., websites offering AI ``nudification'' services. While there may have been other types of applications to generate SNEACI, we specifically look at AI ``undressing'' applications. During this time, we did find one application which was also an AI Face Swapping application; however, the method in which it was identified classified it as an AI ``undressing" application. As this AI face swapping application may likewise be encountered by people seeking AI nudification applications, we kept it in our dataset.
We continued to collect AI nudification applications until we hit saturation, i.e., no new websites appeared with further iterations. 
First, we reached out to a number of NGOs known for their interest in SNEACI
and were given different lists of nudification applications.
Second, we conducted preliminary Google searches to find ``Top [\textit{X}] Nudifying Apps'' articles. These articles outline several AI nudification applications and often discuss features.
After identifying AI nudification applications from the NGOs' lists, we hit saturation of AI nudification applications after visiting three ``Top [\textit{X}]'' articles.
Between these two categories of sources, we found a total of 24 websites.
While we initiated our study and developed our methodology much earlier, at the time of our final walkthroughs and data collection for this paper (July 2024), four of these websites were no longer accessible. As such, our researchers examined 20 website applications hosting AI nudification tools. While these websites would be easy for others to find, we choose to not name the websites in this paper because we do not believe that it would be responsible for driving more traffic to them.

\subsection{Application Walkthrough}
\label{sec:methods:walkthrough}
Light et al.~\cite{light2018walkthrough} lays out a walkthrough of application UI's that address the app in three stages: {\it entry}, {\it everyday use}, and {\it exit}. We use this methodology in our research to complete a critical analysis of AI nudification applications. The walkthrough analysis provides an understanding of the interactions on the website and a user's likely expectations when interacting with the website. As such, throughout this analysis and under the Light et al.~\cite{light2018walkthrough} approach, researchers should take screenshots of user interfaces and make notes on the provided features and how they are presented allowing the researchers to create the complete context of the infrastructure of an application. 

We conducted a walkthrough of the 20 website applications hosting AI nudification tools,
as identified in Section~\ref{sec:methodology:dataset}. These applications present a commercial storefront to purchase the applications and image generation with varying features. We completed a walkthrough of all 20 website applications in two parts. As an addition to Light et al.'s methodology, we double coded the first half of the repository. In the first part, two researchers 
walked through the first 10 websites separately, and we computed a Cohen's Kappa for inter-rater reliability of 0.88 looking at our codes for how the applications present themselves both visually and textually, when age confirmation arises, the app's features, payment processors, and monetization methods. 
Second, to ensure uniformity, one of the researchers from part one performed the walkthrough analysis for the remaining ten websites. The main researcher took screenshots of many of the user experiences upon entry and traversing the site. They also took notes. These screenshots and notes were shared with all of the other researchers for context about the applications and were discussed and refined over the course of numerous meetings.

\subsubsection{Entry}
The initial stage of the Light et al.~\cite{light2018walkthrough} methodology,
\textit{entry}, refers to what the user first encounters as they access an application and extends into account registration and terms of services. In particular, researchers document the sign-up and registration process through well-articulated observation notes and examine the first elements the user interacts with.
This includes required user information for sign-up (e.g., confirmations and questions asked), the registration and sign-in process (e.g., Google sign-in or other sign-in options), and design elements of the entry pages.
The {\it entry} defines the user's experience and affects the user's perception and understanding of the product before 
the user engages the product. As Light et al~\cite{light2018walkthrough} argue, researchers can leverage this analysis to understand the intended use case of the platform.

\textit{In our approach:} At the {\it entry} into a website, each researcher cleared their cookies and loaded the website in an incognito browser for coding and viewing. We denote the features on the {\it entry}, such as the presence of nude images, age verification, sign-in/sign-up, slogans, etc. 

\subsubsection{Everyday Use}
After the \textit{entry} phase, researchers 
move into the \textit{everyday use} stage. 
The {\it everyday use} stage explores 
the functionality of the application and can vary on the extent and time requirements (e.g., basic walkthrough requiring minimal time or a thorough analysis requiring extensive time). 
While full analysis in this stage can provide a thorough understanding of the application, even walking through the basics of an app's functionality can provide an understanding of what activities it enables, limits, and guides the user towards. 
Using this approach, the researchers explore 
not only the available features but also the workflow of using the product (e.g., the various interfaces and how they interact). 

\textit{In our approach:} To map the {\it everyday use} of the website, we first traverse the links within the menu bar at the top of the webpage (e.g., dropdown menu). We collect the terminology used to describe the functionality of the website (e.g., taglines or slogans) that
appears around the website. Although some of these AI Nudification Applications
lock some or all uses of their platform behind a subscription paywall, they report
the functionalities associated with each subscription tier. 
We collected the reported functionality from these
descriptions and the associated cost. The advertised functionality indicates user expectations from the nudification platform. As such,
it denotes the use case, priorities, and values of the creators of the websites. 
However, we find that the terminology used to discuss the functionality varied between applications. While we noted how these applications advertised themselves, we also coded functionality based on the application's functionality instead of only the term used by the application for the tool. 
For example, a website may describe itself as an ``AI undressing
application'' but offer features that allow 
AI face swapping into explicit videos. 

As discussed in Section~\ref{sec:intro}, we specifically look at the \textit{advertised} features mentioned on the AI nudification applications.
A user purchasing these products will likely expect the features to match what the applications advertise. Furthermore, we presume that the advertised features are what the company strives to be able to provide. 

\subsubsection{Exit}
Finally, the third stage of the walkthrough is the \textit{exit}. This stage identifies how the user can exit the application. This includes but is not limited to logging out of the application as well as deleting their account. 
Particularly, walkthrough analysis classifies the user flows for deleting the account and the user's understanding of their
data after account deletion.

\textit{In our approach: } Finally, we finish the walkthrough at the \textit{exit} phase by looking at  if and how a user could delete the account they had created. 

\subsection{Ethical Considerations}
Conversations on the ethics of researching these websites and the actions we took to study them were central to our meetings.
While our work does not interact with human participants, there were several ethical considerations
when conducting this research. 

First, by researching the features of AI nudification applications, we may develop and communicate knowledge about these AI nudification applications that could be used by bad actors. 
Additionally, if we were to name or link to these sites, driving traffic to them could increase their ranking within search engine results. 
As such, we do not name the websites in our study when presenting our results. And, overall, we believe that the knowledge we develop and communicate will also be valuable for those seeking to mitigate the harms of SNEACI.

Whenever we experimented with accounts, e.g., sign-in with Google, we ensured that those accounts were not linkable to the researcher. 
While collecting data from a university campus, the researchers ensured that other people in the environment were not exposed to the imagery on the websites in our study.

Furthermore, and as we would encourage all research efforts that might expose researchers to potentially problematic and triggering content,
we hired a therapist with open office hours every other week that our researchers could visit; the therapist was also available on demand, if needed. 

The contents of this paper may be triggering to some, for a wide variety of reasons. In addition to including a content warning at the start of this paper, we are also hiring an expert for a sensitivity read of this paper, and will revise as appropriate. For example, in addition to thoughtful considerations of how to communicate about image-based sexual abuse, we believe that it is imperative to speak thoughtfully and respectfully about the differences between gender and sex even though the nudification applications themselves do not appear to make such a distinction. Whereas our team feels confident in considering and writing to these types of considerations, we believe it is valuable to hire an external expert for an independent sensitivity evaluation.
In this paper, we chose to show artists renditions of the UI's presented on the AI nudification websites. We believe that the visual depiction more clearly articulates the features and viscerally expresses the harm which was think is important. We chose to do stylized sketches in order to balance reader comfort, expression of research results, and risks of displaying actual bodies from the website.

\subsection{Author Positionality}
As researchers, while we strive to be objective, we realize that our personal backgrounds and perspectives could influence our interpretation of the data that we collected. Thus, we describe relevant elements of our identities and positionality here.
Half of our authors identify as woman, which is the demographic that we found was the primary target of these applications. 

All members of the team believe that the non-consensual uses of these applications are \textit{not} acceptable, which is consistent with the dominant view of acceptability~\cite{BrighamNCEI, umbachNCEI}. Thus, in interpreting our results, we focus on the harms that these systems can create and the rights that are violated when used non-consensually.

\subsection{Limitations}
We would like to acknowledge a number of different limitations; however, we do not believe that they take away from our analysis.
\paragraph{We acknowledge that our repository may not include every AI nudification application; however, we are only trying to characterize the ecosystem.} This case study can be a representation of the population to build upon. 
\paragraph{We acknowledge that the location from where we access these applications may affect what appears on their UI.}
In both creating our database and analyzing the ecosystem, we accessed all of these websites from the United States of America. Other advertised features may be present if accessed from other regions.

\paragraph{We acknowledge that our characterization of the ecosystem is based off of the advertised features.} These features could differ from the actual features present when generating the images through their UI. However, any user accessing the website will believe these are the features as well. Furthermore, we believe as the AI features advance, the applications will get closer and closer to the advertised features. 

\subsection{Terminology}
Throughout the paper we use three terms frequently: ``women'', ``application'', and ``feature''. We use the term ``women'' to describe those depicted on the websites we analyze who appear to the researchers to present as women based on their dress or anatomy (visible breasts, vulva). We acknowledge that there are significant pitfalls in attempting to perceive gender based on visual depiction: gender is a complex concept and typically people self-identify their gender~\cite{scheuerman2020hci}. We use the term ``application'' to refer to the websites that serve as virtual UI store fronts to sell AI tools to consumers to automate the process where an image subject in the submitted image is returned in a ``nude" form. Finally, we use the term ``feature'' to refer to specific features offered by these ``applications'' such as those listed in Table~\ref{table:likeness-based-capabilities}.

\section{Entry: User Experience}
\label{sec:user}

This section reports on the first stage of the walkthrough, the entry phase. Here we study what users encounter as they enter the web applications. Users rarely have to create an account before encountering nude bodies though the option to login/create an account was always present. In this section, we will walk through what users experience upon entering the websites, what was required (or not) of them to be allowed entrance, how and when they had to make an account, and the terms of services of the application.

\textit{Aligned with qualitative methods, our analysis aimed to surface general themes about the AI nudification applications, rather than quantify their prevalence. Accordingly, we report the appearance of themes using the following terminology: a few (less than 25\%), some (25-45\%), about half (45-55\%), most (55-75\%), and almost all (75-100\%).}
% \vspace{-0.1in}
\paragraph{While most websites ask that users are 18 or older, not all do.}

Recall from Section~\ref{sec:methodology} that our data was collected from a region in the U.S.\ that does not have an age verification law for explicit content. 
As our researchers opened each website, in incognito mode and with cleared cookies, seven of the 20 (33\%) websites immediately asked for confirmation that the user is 18 or older. As our team traversed the different pages, 14 out of the 20 \emph{do} ultimately ask the user to confirm their age: seven when the user first loads the website, two when the user goes to login or create an account, and five when users start the process of generating ``undress'' images or purchasing the feature to do so. For the latter two categories (when age verification eventually happens but not at first-load), we find that five  show nude  content prior to confirming the user is over the age of 18; in more detail, three show blurred imagery where nipples and vulvas are blurred~(the rest of the genitals are still visible) and two show complete nudity of the breasts and vulva. 

Of the remaining six~websites that \emph{do not} verify the user is 18 or older throughout the regions of the sites that we navigated, we observe a spectrum of content visible to users. At the extreme, on its landing page, one of these six~websites shows AI-modified images of celebrities engaged in sexual acts along with false news articles about their actions. In contrast, another of these six~websites provides descriptive text of the form ``fake nudes of a person using AI'' but no images. 
While we have not used these six services to create images, as discussed in Section~\ref{sec:methodology}, our study of all six sites suggests that users will be able to generate ``undressed'' images without any age confirmation. 

Additionally, in our walkthrough, we found that only seven of the 20 applications required the image subject to be over the age of 18, and all seven of those only mentioned that requirement in their terms of service. 

\begin{tcolorbox}
\textbf{Finding \#1.}
For some sites, users can \emph{generate} undress images without confirming that they are 18 years old or older. Further, only some sites mention that the image subject should be 18 years old or older, and on all those applications, we only encountered such mention in the terms of service.
\end{tcolorbox}

\paragraph{19 of the 20 websites in our study explicitly specialize in the undressing of ``women'' or the nude female form, though the websites present it in different ways.} 

Most of the websites show their focus on women through imagery. 
17 of the 20 websites in our study show images of women in varying levels of nudity on the landing page. 13 of these 17 sites show women \emph{without} clothing but not in sexual positions, and another three show womens' faces and bodies but include mens' body parts as the women and men engage in sexual acts; mens faces are never shown. 
Of the three websites that do not show photos on the homepage, one has a ``gallery'' with women's nudified images on it with the caption ``nudification samples'' and another specifies in text that it is ``specialized in the female form''. 

Additionally, the website text also communicates a focus on women. In their text descriptions, 14 sites exclusively mention women and/or girls, and no other genders: two of 20 sites refer to the subjects exclusively as ``women'' (e.g., ``AI algorithm for generating nudes from photos of clothed women''), five refer to the subjects as exclusively as ``girls'', and seven refer to the subjects as either ``women'' or ``girls''. For example, one site says ``girl of your dreams''. The use of the word ``girl''
to refer to adult women can be interpreted in a number of ways. For example, it can be viewed as demeaning and sexualizing~\cite{guardianGirls}. However, ``girl'' is also used in empowering ways by a number of different subcultures of hyper-femme expressions~\cite{pornstudies}. Further, two sites use the word ``your'' as in ``your girl'' or ``your woman'' or ``your girlfriend'', potentially suggesting the user's possessiveness of the woman subject. 

In summary, 19 of the 20 nudification application web pages convey a specialization in the nudification of women, though they vary where and how that specialization is communicated. This focus aligns with findings from prior work on non-AI facilitated IBSA, which found that over 90\% of content on ``revenge porn websites'' was of people who based on visual presentation appeared to the researchers to be women~\cite{uhl2018examination}.

\begin{tcolorbox}
\textbf{Finding \#2.}
19 of the 20 undress websites in our study explicitly specialize in the undressing of women, though they vary in the explicitness of the images shown on the landing pages. 
\end{tcolorbox}

\paragraph{\textit{Half} of the undress websites in our study say that users must obtain ``consent'' of the image subject before undressing them.}

Our team additionally studied the consent-related text that users might encounter as they navigate the nudification applications as well as consent-related text in the sites' terms of service, which users may not encounter but which are still present on these sites.

Seven of the 20 sites include text during age verification or before image generation that states that the user must have consent of the image subject.

All seven of these sites --- and an additional three sites for a total of 10 --- include text in their Terms of Service that state that a user needs consent from the image subject to upload their image to the AI generator. 

Consequently, 10 out of the 20 applications do not state a requirement in their Terms of Service that the user must have consent of the image subject in order to have their image modified, and we did not encounter consent-related text during the ``normal'' (non-Terms of Service) user flow in the 13 of the 20 applications.

\begin{tcolorbox}
\textbf{Finding \#3.}
Half of the websites mention in some form that they expect the user to have the image subject's consent before using their tools; however, only some websites ask for the user to confirm this assumption.
\end{tcolorbox}

\paragraph{Registering for an account was simple and mainly gated generation or product purchase.}

Our researchers were able to observe and traverse much of the web pages before being required to register for an account. While users immediately can login or register upon entrance to the web pages in most cases, three applications would allot users login credentials after purchase. For those three applications, users were not immediately able to login or register upon entrance to these sites. And in two of these websites, they could not register an account until after purchase (which the researchers did not do). In one application, users were never given logins. Instead, they can purchase a ``key'' which gives them an set number of generations. 

On 10 out of the 20 of the websites, our researchers did not have to create or login until they were attempting to purchase features or generate images. In four applications, no login is ever required for the researcher to begin to generate imagery. 

\paragraph{Many of the websites allowed researchers to register or login using authentication through other applications.}

In addition to traditional email-based account registration and login, we found that users could also login and register accounts via sign-in through
Discord, Google, Twitter, and Apple. 4 websites allowed Discord-based account sign-in. Google was supported by 12 websites. Apple and Twitter were used to support logins to 3 and 1 websites, respectively.

\begin{figure*}
    \centering 
    \includegraphics[width=\textwidth]{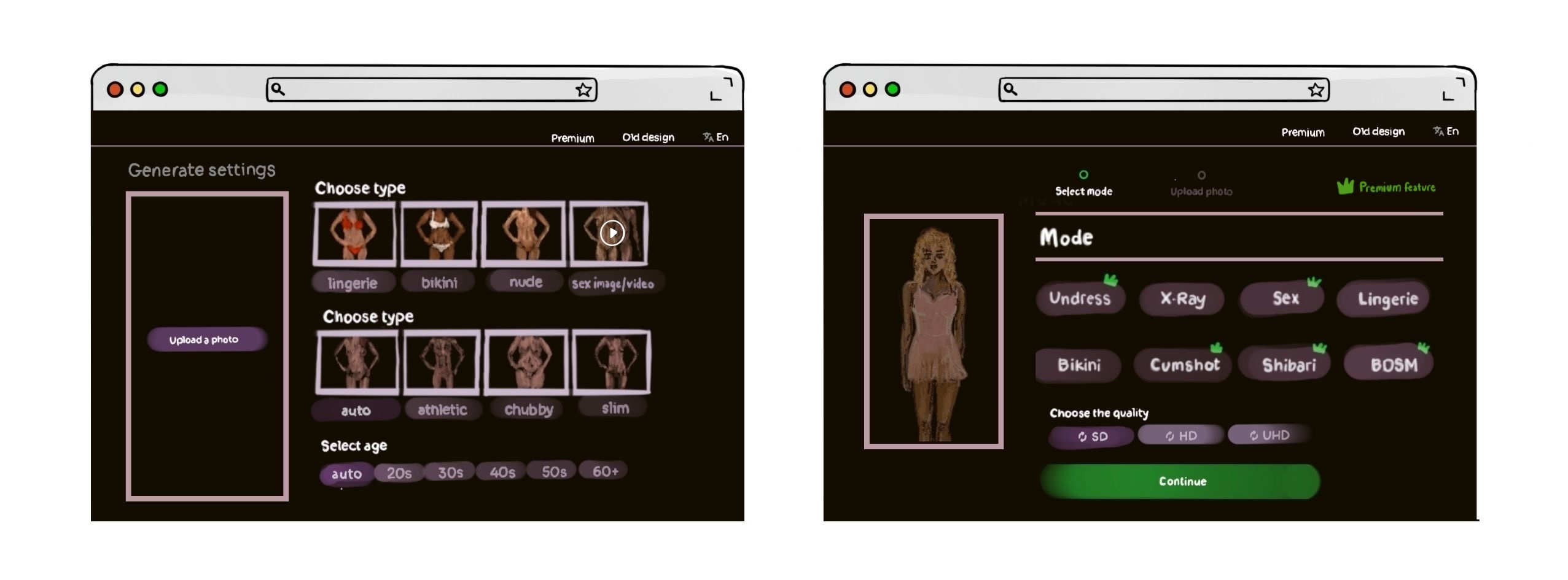} 
    \caption{These are artistic renderings based on several real UI's from the nudification applications we study. In the rendering, less details of the human form are shown. This figure shows the interface for uploading a photo and various different options that a user can pick from to customize the resulting image. On the left, a play button is added over the ``sex'' option to indicate that some applications can produce both image and video outputs. On the right, we indicate that some of the features are only available after payment. Artwork credit: Hanna Barakat.} 
    \label{fig:capabilities-ui} 
\end{figure*}

\section{Everyday Use: Features}
In the second phase of our Light, et al.~\cite{light2018walkthrough}-based walkthrough, we analyze the advertised  features of the 20 nudification applications in our study.

\label{sec:results:features}
\label{sec:results:capabilities}

The nudification ecosystem has a wide variety of features. However on each application there is at least some type of AI feature which allows a user to upload a photo and receive  back a ``nude'' version of the image subject. While the exact feature set may vary, this underlying feature is what defines our repository --- the ability to make someone ``nude'' through an AI feature. Many of these applications allow for users to pay for additional features that allow them to fulfill a number of tasks. This may range from changing what clothing the image subject is in (e.g., putting them into a bikini or in lingerie) to AI-based features that put image targets in AI-altered videos with sexual acts.

\subsection{Likeness-Based Image Features}
\label{sec:results:capabilities:likeness}

We focus our analysis first on the features that these nudification applications offer with respect to the manipulation of the likeness of an image subject (summarized in Table~\ref{table:likeness-based-capabilities}).

\begin{table*}
\caption{AI nudification applications have a wide variety of AI nudification features. Base features such as the AI Undressing Tool, Human-Assisted AI Undressing, and AI Face Swapping are inherently what classify these applications as AI nudification applications; however, some applications offer features that put image-subjects into sexual acts or allow the user to change the image subjects clothing. This table explains each terminology used to denote these features.}
\label{table:likeness-based-capabilities}
\begin{center}
\begin{tabular}{  |p{3cm}||p{11cm}||p{1cm}|  }
 \hline
 \multicolumn{3}{|c|}{\textbf{Likeness-Based Image Features}} \\
 \hline
 \textbf{Feature}& \textbf{Capabilities} & \textbf{Count}  \\ 
 \hline \hline 
 AI Undressing Tool & User inputs a photo of an image subject to receive back a ``nude'' of the image subject. The AI model is essentially trying to predict what is under the clothing. & 18 \\ \hline
 Deepnudes & An instance of an AI Undressing Tool. Users input an image of an image subject. The AI model will modify the image to put the image subject in to different sexual scenarios as specified by the user such as sexual scenes (see rightmost image in Figure~\ref{fig:capabilities-ui}).  & 9\\  \hline
 Clothing Changing & An instance of an AI Undressing Tool. Users input an image of an image subject. The AI model will modify the image to put the image subject in to different clothing such as lingerie or bikinis. & 5\\  \hline 
Human-Assisted AI Undressing & An instance of an AI Undressing Tool. Users give a photo of an image subject to ``Experts'' who use AI tools to make the image subject ``nude''. & 1\\  \hline
\hline
AI Face Swapping & User inputs a photo of an image subject with the option to also input another photo/video of a person and they receive back an image/video where the face of the image subject has been put onto the person of the second image/video. They use default destination photos/videos provided by the service if they do not uploaod one themselves.   
& 1 \\ \hline
\end{tabular}
\end{center}
\end{table*}

\paragraph{There are two broad categories of features that allow a user to see the image subject in the nude: undressing and face swapping. } 

We find that 18 of the 20 applications offers an AI Undressing Tool in which a machine-learning model is purportedly trained to `predict how the breasts and vulva of an image subject look under clothes and alter the image to represent their as-predicted nude bodies (such ML functionality may be termed in-painting or body mesh estimation in the computer vision community~\cite{yu2018generative, luo2020full,huang2022review,bualan2008naked}). 

\begin{figure*}
    \centering
    \includegraphics[height=22em]{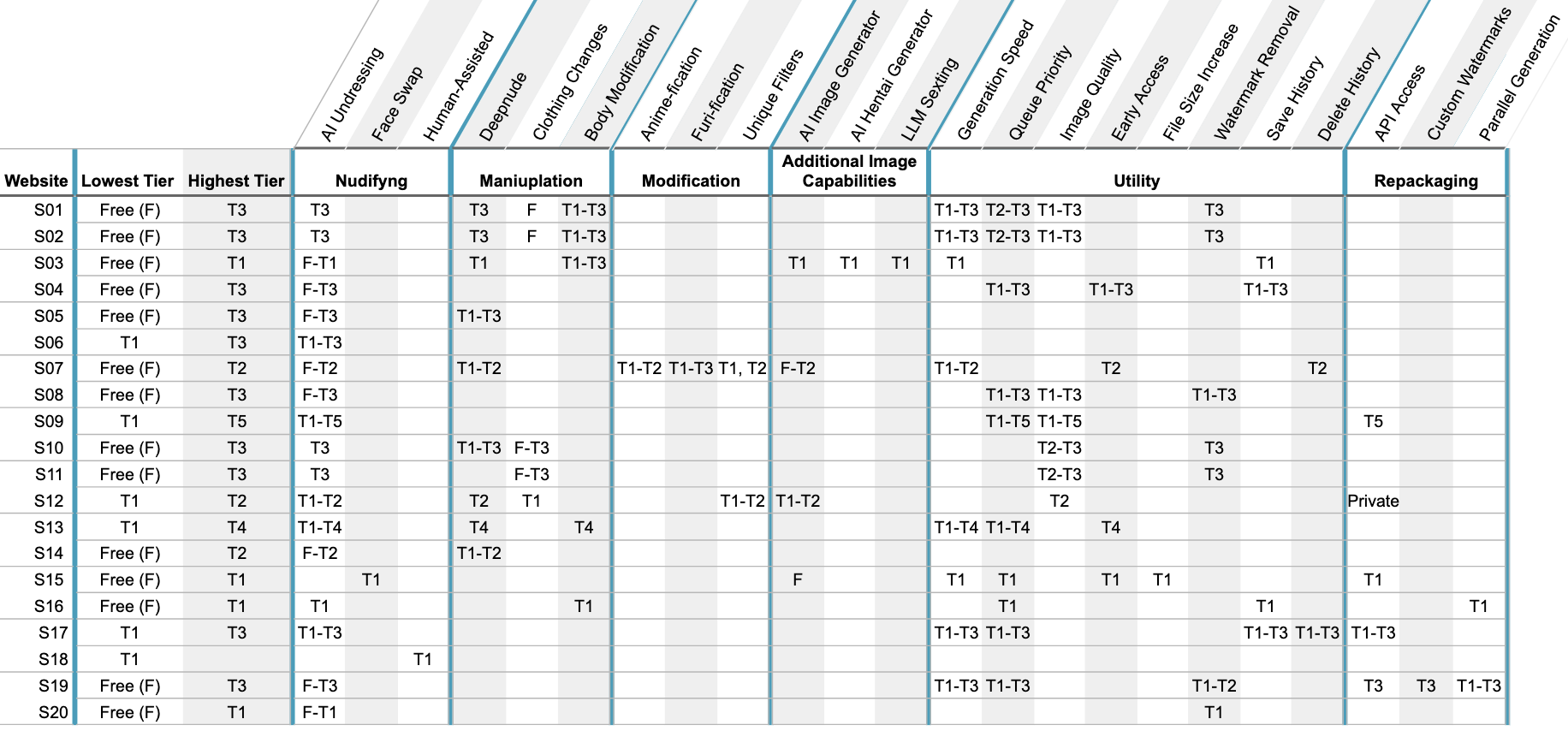}
    \caption{ The ecosystem offers a variety of features at different subscription tiers.
    }
    \label{fig:enter-label}
    \label{fig:spreadsheet}
    % \vspace{-0.1in}
\end{figure*}

Of the two applications that did not offer AI undressing, one application offers Human-Assisted AI Undressing features. Users pay per generation and attributes they want added to the image, but they hire an ``expert'' to do the generation using features that the expert has access to offline. The remaining application offers AI Face Swapping features. This feature essentially presents the image subject nude, but instead of using the machine model to predict how the image subject's body may look under clothing, the AI model is trained to morph the face of someone in an image or video into that of the user's uploaded image subject. In AI Face Swapping, at minimum the user must supply the machine learning model with the image subject's face that will be put onto another person's body; however, they also have the option to supply the materials that the face will be ``swapped'' onto.

It is unsurprising that 18 out of the 20 applications offer AI Undressing tools because our application selection process (see Section~\ref{sec:methodology:dataset}) specifically selected for this feature. However, the two applications that do not advertise AI Undressing but AI Face Swapping and Human-Assisted Undressing were included in the lists of AI Undressing tools we discovered or were provided.
We retained them in our dataset because they are among the applications a user might encounter if they were to seek out AI nudification applications. 

\paragraph{There are also a number of features which allow a user to put a subject in sexual scenes.}
The nudification applications in our dataset do not limit their features to ``undressing'' an image subject in-place. Rather, we encounter a diversity of features when studying the sites in our dataset.

One feature, which we refer to as Deepnudes,  renders an image subject in a 
sexual act. 
See, for example, Figure~\ref{fig:capabilities-ui}, which is an artistic rendering of the UI of one application in our study. In this UI, a user can choose from ``sex''~(e.g., offering ``missionary,'' ``cowgirl,'' or ``reverse cowgirl''), ``cumshot'' (ejaculation on the image subject's face),  ``shibari'' (a form of bondage), and ``BDSM'' (a term for a variety of sexual practices that include bondage, discipline, dominance, submission, and sadism). 9 of the 20 applications in our study offer similar ``Deepnude'' features.  

Adjacent to the above category of features is the Clothing Changes feature, which is present in 5 of 20 applications we study. 
This feature allows a user to apply different clothing to the image subject ranging from bikinis to lingerie. The application that offers AI Face Swapping in images offered this feature for videos as well.

\begin{tcolorbox}
\textbf{Finding \#4.}
Half of applications allow for images to be modified such that the image subject is put into sexual scenes that are not in the original photo.

\end{tcolorbox}
% \vspace{-0.1in}
\paragraph{Some applications allow for modification of the image subject.} Modifications can range from weight, body traits like breast or butt enhancements, or age (18+) changing to even altering the image subject into anime\footnote{A type of art/animation style that is attributed to Japanese animation.} or furi\footnote{A subculture where people dress in fur suits depicting anthropomorphic animal characters.} versions of themselves. Body modifications for weight and age are seen in two and three nudification applications, respectively. Two applications offer unique filters and one application offers both animification and furi-fication models. 

Overall, seven applications offer modes of modifying the images and how the image subject presents in the image.

\subsection{Additional Features}
\label{sec:results:capabilities:other}

We now summarize the additional features advertised on these sites, which speak to the broader set of services offered by these applications and, hence, services that might appeal to some of their users. 

\paragraph{Some applications offer generation features.} 
General image generation is offered by several
of the applications we study. These can be used without any seed image uploaded by the user. Four applications offer general AI intimate image generation, one additional application offers AI Generation of Hentai\footnote{Hentai is an art style of pornographic animated video often attributed to or seen as a subset of Japanese anime.} specifically. 

\paragraph{One application offers synthetic sexual messaging.}
One application offers a trained LLM that will message with users and roleplay different sexual scenes with them. 
While this application's LLM-based sexual messaging features do \emph{not} appear to be connected with its AI nudification services, the fact that a single application offers both services suggests a future in which these services are combined, as hypothesized in prior research~\cite{CheongAIEthics}.

\subsection{Utility}
AI nudification applications offer a number of features to help improve the user's experience that we call ``Utility'' features.

\begin{table*}

\label{table:utility-capabilities}
\begin{center}
\begin{tabular}{  |p{3cm}||p{11cm}||p{1cm}|  }
 \hline
 \multicolumn{3}{|c|}{\textbf{Utility Features}} \\
 \hline
 \textbf{Feature}& \textbf{Capabilities} & \textbf{Count} \\
 \hline
 Generation Speed Increase & Users get increased generation speed for each subscription tier they purchase. & 8\\ \hline
 Queue Priority & User purchase higher subscription tiers to increase their queue priority. & 10 \\ \hline
 Generation Quality Increase & The subscription tier which the user purchases decides how high quality the images they generate can be.& 7\\ \hline 
 Early Access& Users are given access to new features before lower paying users. & 4 \\ \hline
 File Size Increase & Specific to AI Face Swapping, user's can input files with larger file sizes for each subscription tier they pay for.& 1\\ \hline 
 Watermark Removal & At the highest subscription tier, users can remove the watermarks from their images.& 7 \\ \hline 
  Repackaging & Users gain access to an API that can communicate with the AI Model and sell access to it on their own applications.
  & 5\\
 \hline \hline
  Saving History & Users can save images that they have created through the applications features. & 4\\ \hline \hline 
Delete History & Users are able to delete their history of images or videos they have created through the applications features. & 2\\
 \hline
\end{tabular}
\caption{AI nudification applications offer a number of features to help improve the user's experience that we call ``Utility'' features.  
This table defines each term and the frequency with which we observed such features in our dataset.
}
\end{center}
\end{table*}
\paragraph{Enhancements to the generation and photo quality are available.}

Respectively, eight, 10, and seven applications offer increased generation speed, priority queues, and quality improvements through paid offers. We find that four applications also offer early access to future features with paid subscriptions. The application that offers AI Face Swapping also allows an increased file size for the option where paid users input a video for the image subject's face to be swapped into. 

\paragraph{By default, many of these applications watermark their images, and users have to pay for the product's highest tier for them to be taken off.}

Through our application walkthrough, we find that seven of the undressing applications watermark the images they output and sell the removal of those watermarks at the highest tier of their application's subscription.

\paragraph{Some companies allow their AI Model to be repackaged and sold on different applications as their own AI model by giving access to their API. }

\textit{Repackaging} is when a company with a product, in this case the machine learning model for nudifying image subjects or any of the other features listed in this section, sells the product to other companies  with the agreement that the other company can use their own label on the product but have little to no control over the product itself other than the label on it. In business literature, such repackaging is also referred to as ``white labeling''~\cite{whitelabeling}, though we do not use that term here because of its similarity to ``white listing''.
We find that five
 of the 20 applications in our study offer access to their machine-learning model through an API. 
Additionally, one of these applications offers custom watermarks. Two applications explicitly offer parallel generation, where they can have multiple images being generated at the same time, which helps if someone wants to sell the product on their application. 

\begin{tcolorbox}
\textbf{Finding \#5.}
Some applications offer access to their models through an API, allowing other applications to offer the same features without training their own AI nudification tools. 
\end{tcolorbox}
\paragraph{Many applications sell users the ability to save or hide their history.}

Applications are not in agreement on whether a premium feature is the ability to \textit{save} or \textit{delete} a user's history.

We find that two applications make \textit{deleting} a user's history a feature. In contrast, the ability to save past generations is a feature on four applications.

\section{Everyday Use: Monetization}
As part of everyday use we additionally examined the cost of the applications' features, how the applications monetized, the relationships to monetize, and the payment methods that enabled their monetization. 
% \vspace{-0.1in}
\paragraph{All nudification applications are commercial: they \textit{sell} their features and offer very limited free functionality.}

Figure~\ref{fig:spreadsheet} details the features at different paid tiers of the applications we analyze. Fourteen applications offered free features, although the functionality of those features without payment was limited. Eight applications offered ``free'' nudification features, however all of these applications output the ``free'' nudified image in a blurred form or with a large enough watermark so as to incentivize payment to remove the blurring or watermarking. Four additional applications offered clothing changes for free, and two applications offered image generation for free. This was the extent of ``free'' features provided by the applications.

Across all 19 applications that offered AI undressing at scale, i.e., not the human-assisted AI undressing application, and if we consider the lowest-cost paid tier, the mean price per image was \$0.31 (min: \$0.06, max: \$1.00). At the mean price, the cost to generate 1000 such images would thus be \$310, but at it's lowest pricing, that could be only \$6.00.

\begin{tcolorbox}
\textbf{Finding \#6.}
All AI nudification applications are commercial platforms, which could be a vector to incentivize guardrails against abusive use.
\end{tcolorbox}
% \vspace{-0.1in}
\paragraph{Platforms offer a symbiotic relationship with their users to give them a stake in the success of the platform. }

They do this in three relationships: Platform to User, Platform to Platform, and User to User(s). While traditionally we see monetization as a company selling either their product to other companies to sell~(i.e., platform to platform) or company to consumer~(i.e., platform to user), we see the rise a new third monetization relationship that essentially makes certain users a stakeholder in the company's success~(i.e., user to users)~\cite{tradMon,affilmarket}. In fact, eight of the  AI nudification platforms in our study
offer an affiliate program where select
users make a percentage of the revenue generated by new users who create an account on the platform through their affiliate link. 

In the nudification ecosystem we have seen affiliates being offered a range of 25\% to 50\% of the revenue. One application offered 25\%, two offered 30\%, two offered 40\% and one offered 50\%. Furthermore, two applications hid the split that they offer to their affiliates.

Because of the selective nature of these programs, our researchers could not identify all of the forms which an affiliate's pay could be paid out; however, we were able to identify that three of the eight platforms which publicly discuss it  offer payment in cryptocurrency.  

\begin{tcolorbox}
\textbf{Finding \#7.}
Affiliate programs were offered by some of the applications. These applications offered between 25\% and 50\% of their revenue to their affiliates.
\end{tcolorbox}

Five of the 20 applications offered a referral program to all of its users. All users could \textit{refer} other users to the website. When the new user made an account, the account which referred them gains 2-3 credits that they could use for whichever tier they had already purchased. 

\begin{tcolorbox}
\textbf{Finding \#8.}
Referral programs were offered to all users by some of the AI nudification applications. Referrals gave users 2-3 generations at their subscription level.
\end{tcolorbox}
\paragraph{Traditional monetization relationships between the platform and users or to other platforms can be broken into three subgroups.}

Our research found there are 3 different monetization schemes used by the applications: \textit{Feature Scaling } subscriptions, where the higher tier subscription a user pays for the more features they are given access to, normally with undressing and removal of watermarks at the highest tier and the quality, speed of generation, queue priority and number of generations allotted scaling with each tier; \textit{Pro-Credit} subscriptions, where the higher tier subscription a user pays for gives a higher base number of generations allotted as well as a higher discount from which they can purchase further generations but all of the features are \textit{generally} the same; and \textit{All-for-One} subscriptions where one price purchases all features of the product. 

Eight of the 20 applications we study use the \textit{Feature Scaling } subscription model. On average, these applications cost \$0.34 per generation of image over all tiers. 
%\todo{Is previous sentence on average over all these applications \emph{and} over all tiers? or just Tier 1?}
As a general trend, purchasing a higher tier of subscription does decrease the price per generation; however, it is not a one-to-one because higher tiers allow the user access to a larger feature set. For example, as seen in Figure~\ref{fig:spreadsheet}, Application S01 allows for Clothing Changes as a "free" tier, but Undressing is not accessible under Tier 3 and all images will be Watermarked till Tier 3 where the user gains access to Watermark Removal. 

Seven out of the 20 web applications used the \textit{Pro-Credit} subscription model. While AI nudification features are constant between the different tiers, this subscription also purchases scaling discounts for future credit purchases which need to be taken into account when looking at the cost-per-generation. On average, the cost-per-generation over all tiers for these applications is \$0.30 which is slightly less than the Feature Scaling subscription model.

The \textit{All-for-One } subscription model was used by four web platforms. These applications only ever had one price. On average, they cost \$16 per month. Users could generate each image for an average \$0.35. 

The one application that offered Human-Assisted AI nudification made users pay at a rate of \$20 per undressing image, and more for an image with greater sexual content. This is an outlier for our dataset and represents something closer to a crowdworker paradigm than a point-and-click application paradigm.

\paragraph{Five of the applications offer APIs for service resale.}
Five of the applications offered repackaging options where users could gain access to their model through an API.  The cost of API access ranged from \$20 to \$299 with a mean of \$92. At the highest tier of the subscription, one platform offered repackaging options for \$299 per month. This platform offered those who purchase that tier to ``resell our service using API''. This tier also gave access to a manager and marketing plan.

\paragraph{Cryptocurrency is the most popular avenue for payment in this ecosystem. }
The commercial aspect of these platforms means that all applications must have some way to get their payments. Many of these applications rely on third party payment processors or cryptocurrency transactions. 

Almost all of the nudification platforms heavily rely upon or incentivize 
paying in cryptocurrency.
17 out of the 20 applications offer users the option to pay through cryptocurrency. Of those, four of the platforms use incentives to encourage users to pay with cryptocurrency. They do this by offering an additional 10-20\% discount on the subscription purchase. Furthermore, five of the 17 platforms offer cryptocurrency as their only payment method. 

Of the three web platforms that did not offer cryptocurrency payment capabilities, based on their UIs, these platforms relied on PayPal and credit card payments to generate their revenue.

Eight of the 20 platforms offered the option of PayPal payment while seven accepted credit cards.
Two nudification platforms offered to be paid through Patreon and three offered Apple Wallet.
We also found that Steamskins, Venmo, PhonePe, Alipay, CashApp, Google Wallet, and Amazon Pay were each accepted by one platform. We include a figure representing a visual breakdown of these results in Appendix~\ref{fig:payment-processors}.

\section{Exit: Account Deletion}

At the third stage of an application walkthrough, we look at the ease of account deletion. For this step, we made 12 accounts through Google's single sign on when it was offered. On the other five websites, we created them through email. The final three websites did not allow us to make accounts.
\paragraph{Only one application let our researcher's delete their account through the website.} Our researcher made accounts on 17 of the 20~(three websites did not allow for account registration) websites and looked to see how to delete their account. Only one of these websites allowed the researcher to delete their account.

\section{Discussion and Conclusions}
\label{sec:discussion}
Using the walkthrough methodology of Light et al.~\cite{light2018walkthrough}, we study and develop an informed understanding of the nudification application ecosystem on the web. We find that point-and-click AI nudification features are ubiquitious, that these applications position themselves to specialize in the generation of nude or sexually explicit images of women. Below, we reflect upon our findings, first by considering our findings in the context of the impact to potential victims and the role of consent (or the lack thereof), and then by exploring approaches to mitigate SNEACI.

\subsection{False Representations of One's Body}
Representations of self are profoundly personal regardless of their form: static images, textual representations of self in early online communities through emerging representations like VR avatars.  

Prior research has found that digital self-representation~(e.g., avatars) can have significant effects on one's self-concept, and on social behavior~\cite{avatarVio}. Since the MOO communities of the 90s, people have described experiencing harms to their digital representations much like harms to their physical bodies~\cite{mackinnon1997virtual,citron2018sexual}. Prior work on non-synthetic IBSA confirms this finding, noting that the ``experience of image-based sexual abuse [is] embodied --- experienced in and through their bodies, altering their sense of bodily integrity''~\cite{henry2020image}. Such harms are not merely ``emotional'', but pervasive throughout victim-survivors' lives. Prior work notes for example that ``companies may refuse to interview or hire women and minorities because their search results include nude images or deep-fake sex videos''~\cite{citron2018sexual}. Even consensually-altered pictures of one's self can have negative effects on people's self-esteem and lead to self-objectification~\cite{alterVio}. As Citron writes in ``Sexual Privacy''~\cite{citron2018sexual}, ``even though deep-fake sex videos do not depict featured individuals' actual genitals, breasts, buttocks, and anuses, they hijack people's sexual and intimate identities.''

Historically, non-consensual photos and their sexual alterations were mainly through cutting out faces from magazines or public photos and tapping them to intimate imagery~\cite{photoCutting}. This type of alteration still meant that image subjects had to have their images publicly printed, and the altercations were relatively noticeable. After that, we saw a rise in the number of photoshopped images~\cite{Hoggard_2022}. However, their resemblance to the original was mainly gated by the user's skill level in Photoshop. Now, with the rise of commercial AI Nudification platforms, users without technical skills may still create altered images of a person, and they can do so with a few clicks of a  button. 

Our analysis finds that existing AI nudification tools appear positioned to specifically create abuse material of those who have not consented to their images being used. Below we offer suggestions for reducing the availability of these abusive tools. 

We note that it is possible that there are legitimate sexual expression uses for such tools. 
Enabling potential non-abusive use cases while not allowing for abuse will hinge on verification of consent of the individual who is depicted in the images uploaded into such tools. This was not the case in the applications we studied. As seen in Section~\ref{sec:user}, only 10 of the 20 applications mention consent in their ToS. 
Applications aiming to offer consensual nudification functionality could draw on existing practices in the sex work industry to verify consent of the individual depicted in the input imagery, and that the individual depicted is over 18, by using a combination of: 2257 forms~\cite{2257cert}, which document two forms of ID to verify that performers are over 18, manual review of every piece of uploaded content as done by PornHub~\cite{PornHub}, and cross-checking of identity verification via biometrics with those depicted in the uploaded content on a recurring basis, as done by  OnlyFans~\cite{OnlyFans}. We note that centralized approaches have inherent privacy tradeoffs, as they require the company verifying consent to maintain a collection of sensitive identity information~\cite{stardust2023high}. 
We found no instances of such verifications in our study of 20 nudification applications. 

\subsection{Reducing the Availability of Abuse Tools}
Prior work in computer security research suggests the value of considering the full pipeline of actions and actors in an abusive ecosystem~\cite{coalitionAgainst}, and then 
 advocating that those stakeholders supporting elements of the abusive ecosystem remove their support (e.g., demonetization). Such advocacy requires (a) identifying stakeholders in the pipeline and (b) respecting freedom of expression, including sexual expression, by distinguishing abusive website behavior from non-abusive behavior such that actions are restricted to abusive websites. 

\paragraph{Stakeholder Identification.}
Our research identifies the existing stakeholders in the nudification pipeline as: the creators of the models used to power the nudification websites, the nudification websites themselves, the payment processors for the nudification websites, single-sign on providers for the websites, 
and platforms that support discovery of the websites through search or app stores. We identify and catalogue these stakeholders via manual analysis. Future work may seek to build an automated stakeholder identification pipeline. This could be a difficult process; however, our research by characterizing the ecosystem begins the work of identifying stakeholders such as payment processors that enable the commercialization of AI notification applications. 

Such information could be used by regulatory bodies or voluntarily by platforms that provide single-sign on or application discovery 
to reduce their contribution to the abuse ecosystem. 

There may be technical challenges to identifying some stakeholders. Many of the nudification websites we analyze refer to having been built on top of other open-source models. Indeed, the functionality used by these platforms has long been a subject of study in the computer vision community (e.g., inpainting, body mesh estimation). Consequently, there is a long history of open-sourcing models that offer the functionality necessary to launch an AI nudification website, without requiring the users of those models to have significant expertise. The proliferation of open-source models on platforms such as Hugging Face in recent years have added to the availability and ease of offering such functionality at commercial scale. 

While the computer science community has long centered the importance of open-source software and data,
recently the ML community has shifted toward considering whether access to high risk research tools should be restricted: ``Models that have a high risk for misuse or dual-use should be released with necessary safeguards to allow for controlled use of the model, e.g., by requiring that users adhere to a code of conduct to access the model''~\cite{neurips2023}. Such steps have precedence in other domains such as medicine: NIH for example requires an application to access genetic data they hold in their database~\cite{nihGenetic} and computer science community projects like the National Internet Observatory similarly require an application and multiple board reviews before data / tool access are provided~\cite{internetObserve}. 

Future work should consider what steps for responsible use should be taken by open-source model providers, including but limited to tracking downstream uses of their models for abuse and governing access to model features, as well as technical interventions for identifying ``responsible'' model providers whose features underlie downstream abusive websites or copycat models.

Additionally, our research identified that five of the applications we analyzed offered access to their functionality via an API such that a user could repackage the application as their own on on another website. This option to repackage the AI nudification application onto another store front brings up a number of concerns. For one, while these commercial sites already decrease the cost of entry for users to create abusive content, repackaging these applications also decreases the cost of entry for new applications building on existing functionality. 
Whether the initial application means for their AI nudification features to be used in non-consensual use cases or not, they do not have control over how the new application decides to market these AI nudification features. 
Furthermore, 
repackaging could act as a shield for malicious actors allowing for a chain of shell companies to hide behind.

\paragraph{Signals to Identify Abusive Behavior.} There are a number of signals that could be used to identify abusive nudification websites, with the goal of distinguishing them from (a) general GenAI websites and (b) non-abusive GenAI focused on sexual expression: e.g., the text used on the website, the SEO terms and/or advertisements used to promote it, and the prevalence of use of the site for abusive behavior. 

Our work finds that the lexicon used by these websites to promote their services is still evolving, and varied, with sites describing the same features in several different ways. However, a common lexicon begins to emerge, particularly around the promoted use cases of the services with e.g., the use of phrases like ``AI Undressing'' or ``AI Deepnude.'' A common lexicon of terms that arise with high frequency in the text or SEO keywords of abusive websites and distinguishably lower frequencies on non-abusive websites could allow for automated identification of those sites to, e.g., payment processors, single sign-on providers, model owners whose models are being used by those platforms. 

Future work may seek to map the pipeline of clearly abuse-suggestive advertisements that link back to particular websites. We hypothesize based on emerging evidence from investigative journalism~\cite{Maiberg_2024, Maiberg_2024a} that the advertisements used by abusive sites may be clearly distinguishable from those used by non-abusive sites.

Finally, our research finds that some AI
nudification platforms already watermark the content they produce.
Regulators with a centralized processing infrastructure for reports of abusive imagery ~\cite{eSafety} could use these watermarks for attribution: to keep track of which websites are generating the most reported abuse material. There are two current limitations to this approach. First, our analysis found that several of the AI nudification websites we examined offered the option for users to pay to remove the watermark applied by the platform. 
Regulation to require AI-generators to watermark their content is being considered in multiple jurisdictions~\cite{provenanceBill} and would be necessary to give regulatory enforcement power against platforms that do not comply. Second, regardless of any regulations, 
watermarks would need to be resilient to removal by users or others~\cite{Wu2024, surveyPetit}.

\section*{Acknowledgements}

This work was supported by the US National Science Foundation under grants CNS-2205171 and CNS-2206950. We would also like to acknowledge the work done by Hanna Barakat to help us with our figures, and the sensitivity reads completed by Miranda Wei and Vaughn Hamilton. 

\bibliographystyle{abbrv}
\bibliography{main}

\appendix

\section{Appendix A: Age Confirmation}
These applications create sexual content that may not be appropriate for younger ages. During Section~\ref{sec:user} we discuss when a user is asked by the application to confirm their age. To visually represent this, we show a flow diagram of when all 20 applications ask a user to confirm their age. This "confirmation" is only a button that they have to click to continue through the website.

\begin{figure}
    \centering
    \includegraphics[width=1\linewidth]{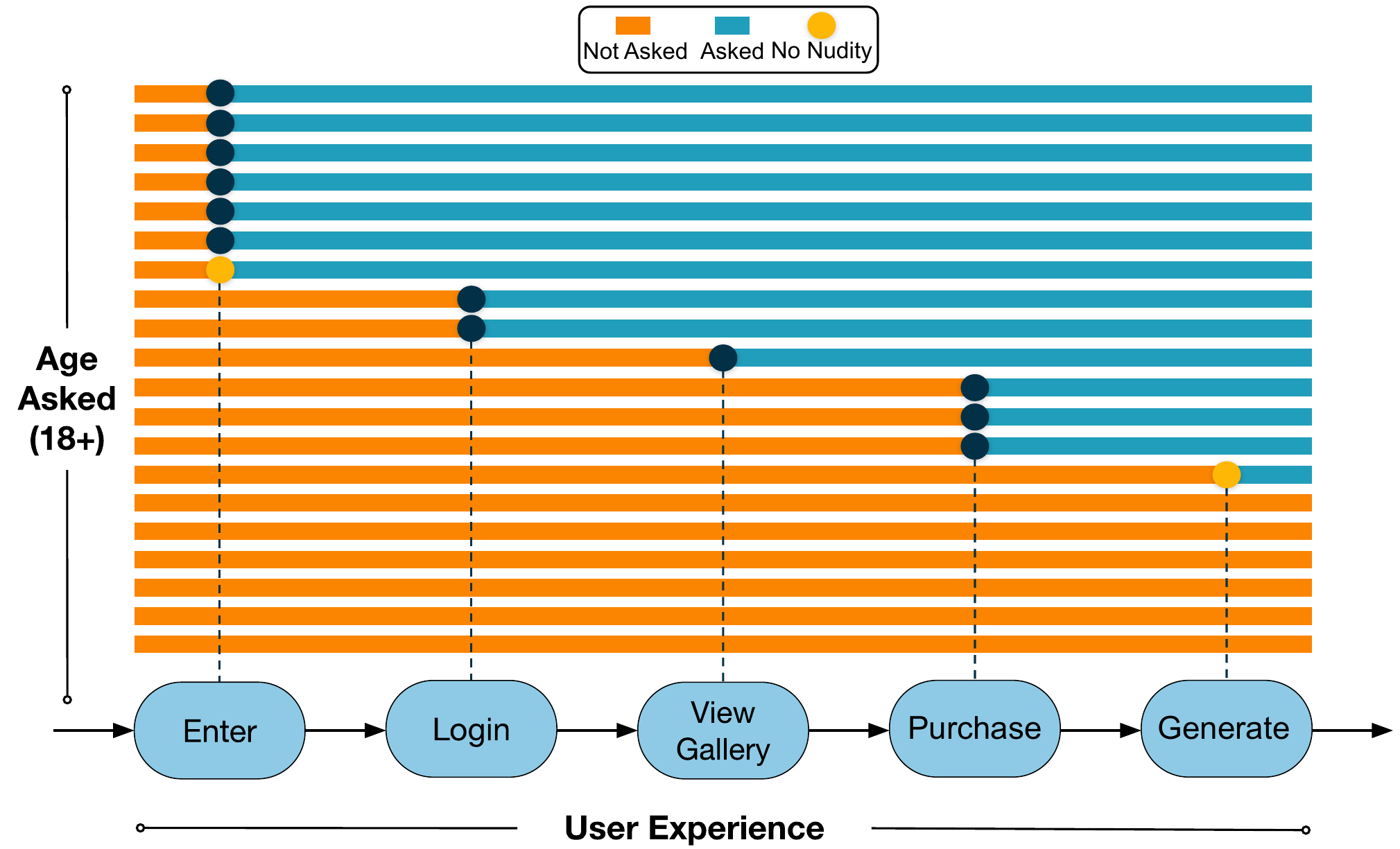}
    \caption{14 out of the 20 websites will ask the user if they are 18 years old or older; however, when this occurs varies. 7 websites ask upon loading up the website; 2 ask when the user registers/logs in; 1 asks when users attempt to view the gallery; 3 ask before purchase; and 1 asks at generation. These periods reveal a variety of nudity to the user before their age is asked, and in the case of 6 applications, they are able to generate images without being asked their age.
    }
    \label{fig:enter-label}
\end{figure}

\section{Appendix B: Imagery Present}
In Section~\ref{sec:user}, we discuss the types of imagery the user may come across during their entry walkthrough: women's bodies both clothed, nude, in sexual acts. However, the further flesh out the spread of imagery, we created a bar graph to show what a user may view on the homepage.
\begin{figure}
    \centering
    \includegraphics[width=1\linewidth]{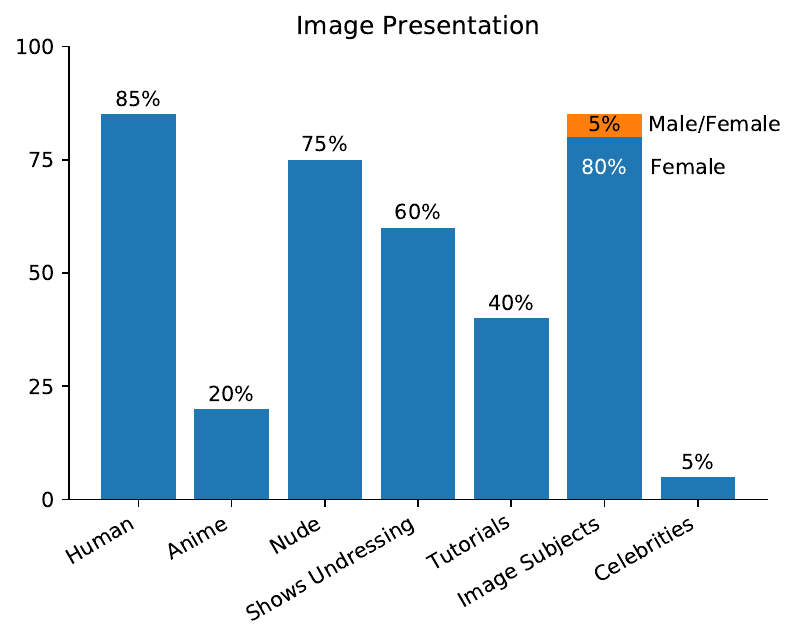}
    \caption{All 20 websites specialize in AI Undressing Tools for the female form; however, the levels of explicit imagery visible on the homepage of each application 
    vary between websites. This can range from blurred naked ``undressing" photos to celebrities being posed naked in sexual positions.
    }
    \label{fig:enter-label}
\end{figure}

\section{Appendix C: Pricing breakdown}
To further discuss and lend intuition into how these applications are being funded, users can visually see how the payment processor breakdown looks, and the prices per tier that users pay as well as the average price-per-image on each application.

\begin{figure*}
    \centering
    \includegraphics[height=25em]{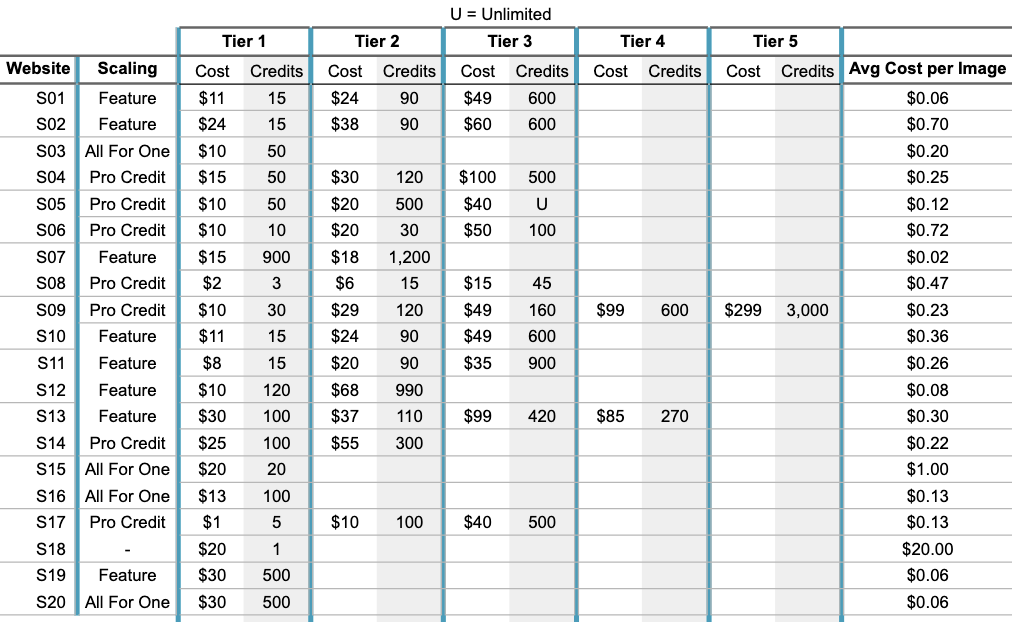}
    \caption{ This ecosystem offers a variety of price points. On average, AI nudification generations cost between \$1.00 and \$0.06 depending on the application that the user bought from and the subscription tier that they purchased.
    }
    \label{fig:enter-label}
\end{figure*}

\begin{figure}
    \centering
    \includegraphics[width=1\linewidth]{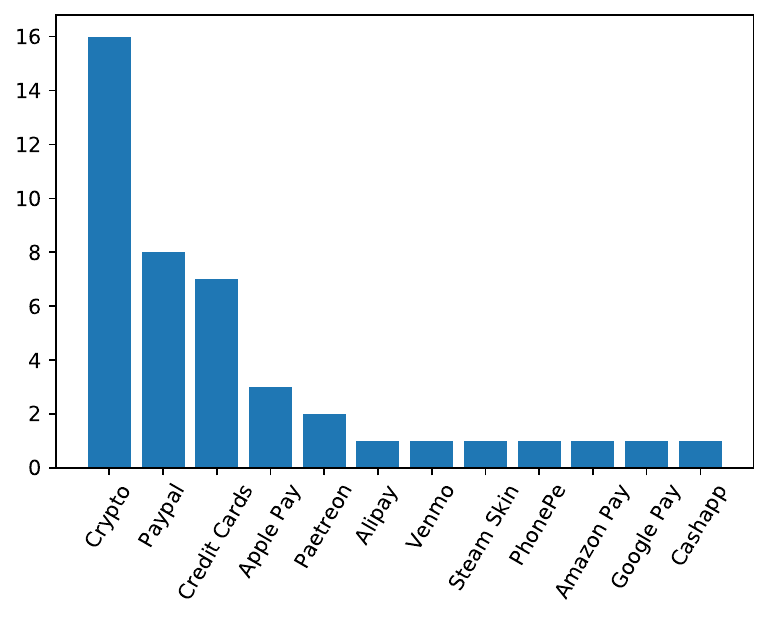}
    \caption{AI Nudification Platforms offer a number of payment processors to complete transactions with them and allow users to use their AI Nudification Tools. The most popular is cyptocurrency, followed by paypal and creditcards. However, some websites allow for apple pay, paetron, alipay, venmo, stream skins, phonepe, amazon pay, google pay, or cash app to complete the transaction.
    }
    \label{fig:payment-processors}
\end{figure}

\clearpage

\end{document}